\input psfig
\documentstyle[twoside,fleqn,espcrc2]{article}

\newcommand{\AmS}{{\protect\the\textfont2
  A\kern-.1667em\lower.5ex\hbox{M}\kern-.125emS}}

\title{Recent results using all-point quark propagators}
\author{ A.~Duncan\address{Dept. of Physics and Astronomy,University of Pittsburgh,
 Pittsburgh, PA 15260},%
  E.~Eichten\address{ Theory Group, Fermilab, PO Box 500, Batavia, IL60510},%
  and 
  J.~Yoo $^a$ %
  \thanks{Talk presented by J. Yoo}}
\begin{document}
\begin{abstract}
Pseudofermion methods for extracting all-point quark propagators are reviewed,
with special emphasis on techniques for reducing or eliminating
autocorrelations induced by low eigenmodes of the quark Dirac operator.
Recent applications, including high statistics evaluations of hadronic
current correlators and the pion form factor, are also described.
\end{abstract}

%%%%%%%%%%%%%%%%%%%%%%%%%%%%%%%%%%%%%
%%%%%%%%%%%%%%%%%%%%%%%%%%%%%%%%%%%%%
\maketitle
%%%%%%%%%%%%%%%%%%%%%%%%%%%%%%%%%%%%%
%%%%%%%%%%%%%%%%%%%%%%%%%%%%%%%%%%%%%

\section{Introduction}

 In many cases the extraction of multipoint hadronic correlators of physical
interest requires enormous statistics, which, given the difficulty of 
unquenched simulations in QCD,
makes it critical to extract the maximum physical content of
 each available configuration. Recently the pseudofermion approach \cite{Lat01} to
the calculation of all-point quark propagators has been substantially
improved \cite{DuncEich} by the use of mode-shifting, effectively eliminating
autocorrelation problems and allowing the accurate extraction of all-point quark
propagators for arbitrary sources and sinks.  In this talk, some applications
of these methods will be described. We will describe results for (a) fits of
current correlators to chiral Lagrangian forms, (b) studies of the pion form-factor
on 10$^3$x20 unquenched lattices (with 200 MeV pions), and (c) the problem of
disconnected parts in isoscalar channels.

\section{Methodology}

 We begin from a bosonic pseudofermion field $\phi_{ma}$ with action ($m$ a lattice site, $a$ the spin-color index, $Q$ the Wilson or clover operator):
\begin{eqnarray*}
   S(\phi)&=& \phi^{\dagger}Q^{\dagger}Q\phi \\
          &=& \phi^{\dagger}H^2 \phi,\;\;\;H \equiv \gamma_{5}Q = H^{\dagger}
\end{eqnarray*}
For fixed background gauge field $A$, simulating the field $\phi$ 
produces the  correlator (where $<<O>>$ means the average of
 $O$ relative to the measure $e^{-S}$)  :
\begin{eqnarray*}
   <<\phi_{ma}\phi^{*}_{nb}>>_{S(\phi)}&=& (H^{-2})_{ma,nb} \\
   <<\phi_{ma}(\phi^{\dagger}H)_{nb}>>_{S(\phi)} &=& (H^{-1})_{ma,nb}\\           
   &=&(Q^{-1}\gamma_{5})_{ma,nb}
\end{eqnarray*}

 Unfortunately, low eigenmodes of $H$ frequently result in very long
autocorrelations for the low momentum parts of hadronic correlators. These correlations
can be removed (or substantially reduced) by shifting the guilty IR modes into the UV:
define a  shifted hermitian Wilson-Dirac operator, with $\lambda_{i},i=1,..N$ the $N$
 lowest eigenvalues, and $\bf{v}_{i}$ corresponding eigenvectors:
\begin{eqnarray*}
 H_{s}&\equiv& H + \sum_{i=1}^{N}\delta_{i}\bf{v}_{i}\bf{v}_{i}^{\dagger},  \\
   \delta_{i} &\equiv& \lambda_{i}^{(s)}-\lambda_{i}  
\end{eqnarray*}

\begin{figure}
\psfig{figure=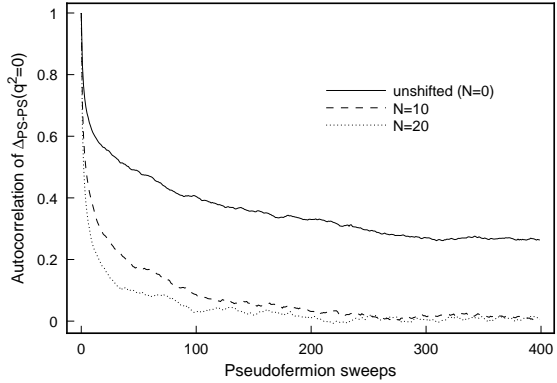,height=0.7\hsize}
\vspace*{-0.4in}
\caption{Reduction of autocorrelation from mode-shifting}    
\end{figure}

Take $\lambda_{i}^{(s)}= \rm{sign}(\lambda_{i})$: this evacuates the IR part
of the quark operator, greatly reducing low momentum autocorrelations (see Fig. 1).
Then, we reconstruct the correct all-point propagator by adding in the contribution from
shifted modes:
\begin{eqnarray*}
 H^{-1}_{a\bf{x},b\bf{y}}&=& <<\phi_{a\bf{x}}\phi_{b\bf{y}}>>-\sum_{i=1}^{N}\Delta_{i}\bf{v}_{i,a\bf{x}}\tilde{\bf{v}}_{i,b\bf{y}}  \\
  \Delta_{i} &\equiv& 1-\frac{1}{\lambda_{i}^2}
\end{eqnarray*}
Accurate mode-shifted all-point propagators can be obtained even for 
{\em very} light
quarks, when conjugate gradient inversion dies. Moreover, the all-point propagator  
represents a great increase in statistics over conventional one-point
propagators:e.g. on a 10$^3$x20 lattice, the all-point is the average of 20000 
(admittedly highly correlated) conventional
propagators (see Fig. 2).
\vspace*{-0.1in}
\begin{figure}
\psfig{figure=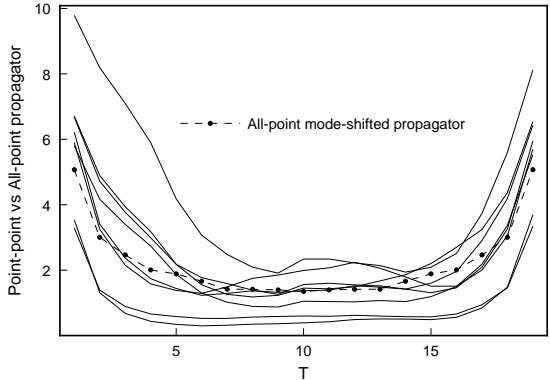,height=0.7\hsize}
\vspace*{-0.5in}
\caption{Comparison of single vs all-point propagators}    
\end{figure}
\section{Applications}
\subsection{Fits of current correlators}
 The large statistics provided by all-point propagators allows a high precision 
calculation of hadronic current correlators, which can then be fit to the predictions
of chiral perturbation theory, ultimately allowing us to extract parameters of the
chiral Lagrangian directly from these correlators. Fits of this sort from pseudoscalar-pseudoscalar
and axial vector-axial vector correlators are shown in Fig.3 (using 800 unquenched configurations
generated on large coarse lattices with the truncated determinant 
algorithm (TDA) \cite{DPY,TDA}).
The values for $M_{\pi}$ and
$F_{\pi}$ extracted from these fits are consistent with those obtained by conventional
fits using smeared-local correlators, but we also obtain an accurate reading of
the size of higher order terms in the chiral expansion.

\begin{figure}
\psfig{figure=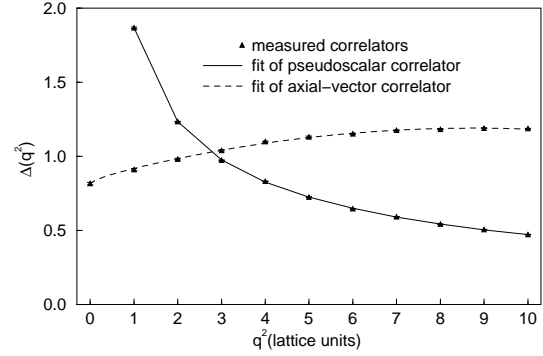,height=0.64\hsize}
\vspace*{-0.4in}
\caption{Fits of pseudoscalar and axial-vector correlators from all-point measurements}    
\end{figure}
From chiral perturbation theory, we have:
\begin{eqnarray*}
  \Delta_{psps}(q^2)&=& \frac{A_1}{q^2+A_2}+A_3+A_4 q^{2}+A_5 (q^{2})^{2}  \\
  \Delta_{axax}(q^2)&=& \frac{q^2F_{\pi}^2}{q^2+M_{\pi}^2}+A_1+A_2 q^{2}+A_3 (q^{2})^{2}
\end{eqnarray*}
The fits to these formulas are shown in Fig.3: for example, for the axial vector 
case we find that the one-loop chiral contribution  $A_{2}$ is extracted with a statistical
accuracy of $<$ 1\%:
\begin{eqnarray*}
    F_{\pi}\hspace{-0.1in}&=&\hspace{-0.1in} 0.187\pm 0.011,  
    A_{1}= 0.8148\pm 0.0038 \\
    A_{2}\hspace{-0.1in}&=&\hspace{-0.1in} 0.0777\pm 0.0003, 
    A_{3}= -0.00442 \pm 0.00002 \\
\end{eqnarray*}
\vspace*{-0.5in}
\subsection{Three-point Functions: the Pion Form Factor}
To extract the pion formfactor, we need the following 3-point function: 
\begin{eqnarray*}
J_{t_{0}t_{1}t_{2}}(\vec{q}^2) &=& \sum_{\vec{w}\vec{x}\vec{y}\vec{z}}e^{i\vec{q}\cdot(\vec{x}-\vec{y})}f^{\rm sm}(\vec{z})f^{\rm sm}(\vec{w})\\
&&\hspace{-1.05in}<\hspace{-0.07in}\bar{\psi}(\vec{z}+\vec{x},t_2)\gamma_5\psi(\vec{x},t_2)j_{0}^{\rm em}(\vec{y},t_1)
\bar{\psi}(\vec{w},t_0)\gamma_5\psi(0,t_0)\hspace{-0.07in}>
\end{eqnarray*}

For large time ($t_0=0 << t_1 << t_2$)  (using optimized smearing functions $f^{\rm sm}$):
\begin{eqnarray*}
J_{t_{1}t_{2}}(\vec{q})\rightarrow (M_{\pi}+E(q))e^{-M_{\pi}t_1-E(q)*t_2}F_{\pi}(q^2)
\end{eqnarray*}
Of course, one needs this convergence before the signal disappears! This requires a large ensemble.
The results for $F_{\pi}(q^2)$ using times $(t_1,t_2)=$ (3,6) and (4,8) are shown in Fig. 4, using mode-shifted all-point
propagators for 65 configurations
of unquenched 10$^3$x20 lattices at $M_{\pi}\simeq$ 200 MeV generated by the TDA method. 
We expect that a useful result for the pion form factor will
emerge from an ensemble of a few hundred configurations, which are presently being generated.
\begin{figure}
\psfig{figure=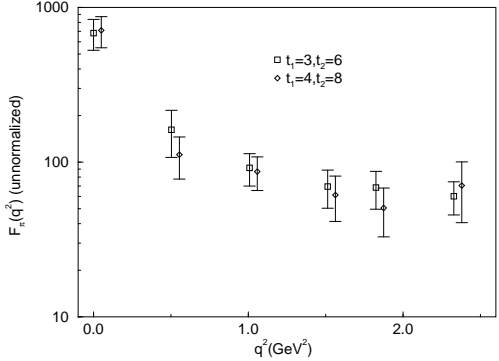,height=0.65\hsize}
\vspace*{-0.3in}
\caption{Connected and Disconnected contributions to Pion Form Factor}
\vspace*{-0.2in}
\end{figure}

\subsection{Disconnected parts and the $\eta^{\prime}$}
 Calculations of the $\eta^{\prime}$ mass from isoscalar correlators suffer from the need
to extract disconnected (annihilation) graphs which require all-point propagators.
Some examples of connected and disconnected contributions to the isoscalar correlator
computed with mode-shifted all-point propagators on 5 configurations of unquenched 6$^4$ lattices
are shown in Fig. 5. The statistical errors of the connected and disconnected contributions
are {\em comparable} using all-point propagators, but the fluctuations are large from one gauge
configuration to the next, requiring a large ensemble before an accurate $\eta^{\prime}$ mass
can be obtained from the subtracted amplitude. However, the difficulties of disconnected
configurations can be avoided entirely by extracting the $\eta^{\prime}$ mass from scalar
isovector correlators which are dominated at large time by a $\eta^{\prime}-\pi$ two-body
s-wave state. The isovector scalar and pseudoscalar correlators extracted from 70 fully
unquenched 6$^4$ lattices ($a$=0.36 F) with two degenerate light sea-quarks ($M_{\pi}\simeq$330 MeV)
are shown in Fig. 6: the corresponding $\eta^{\prime}$ mass is $\simeq$ 735 MeV. 

\begin{figure}
\psfig{figure=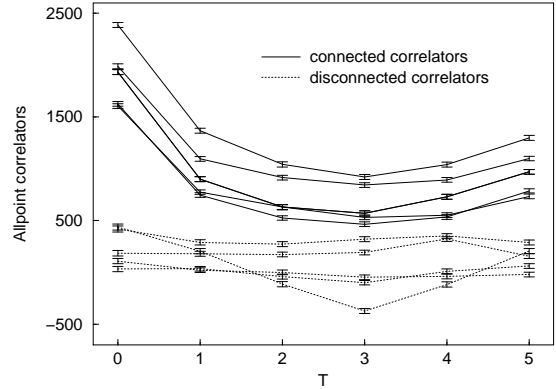,height=0.7\hsize}
\vspace*{-0.4in}
\caption{Connected/Disconnected allpoint correlators}
\vspace*{-0.2in}
\end{figure}

\begin{figure}
\psfig{figure=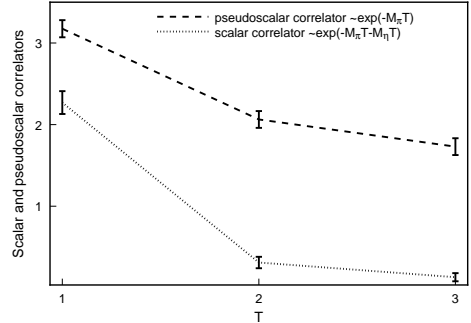,height=0.6\hsize}
\vspace*{-0.4in}
\caption{Scalar and Pseudoscalar Correlators for Unquenched 6$^4$ Lattices}
\vspace*{-0.2in}
\end{figure}

\vspace{-0.1in}
%%%%%%%%%%%%%%%%%%%%%%%%%%%%%%%%%%%%%%
%% REFERENCES %%
%%%%%%%%%%%%%%%%%%%%%%%%%%%%%%%%%%%%%%

%%%%%%%%%%%%%%%%%%%%%%%%%%%%%%%%%%%%%%


\begin{thebibliography}{99}
\bibitem{Lat01} A. Duncan, E. Eichten and J. Yoo, Nucl.Phys.(Proc.Suppl.)B106(2002) 1061.
\bibitem{DuncEich} A. Duncan and E. Eichten, Phys.Rev. D65(2002) 114502.
\bibitem{DPY} A. Duncan, S. Pernice and J. Yoo, Phys.Rev. D65(2002) 094509.
\bibitem{Gass}  J. Gasser, H. Leutwyler, Ann. Phys. 158,142(1984).
\bibitem{TDA} A. Duncan, E. Eichten, and H. Thacker, Phys. Rev. D59, 014505 (1998).
\end{thebibliography}
\end{document}